\documentstyle[sprocl,epsf]{article}

\def\Journal#1#2#3#4{{#1} {\bf #2}, #3 (#4)}

\def\NPB{{\em Nucl. Phys.} B}
\def\PLB{{\em Phys. Lett.}  B}
\def\PRL{\em Phys. Rev. Lett.}
\def\PRD{{\em Phys. Rev.} D}

\def\be{\begin{equation}}
\def\ee{\end{equation}}
\def\bea{\begin{eqnarray}}
\def\eea{\end{eqnarray}}

\begin{document}

\begin{flushright}
\begin{tabular}{l}
  SLAC-PUB-7173\\
  hep-ph/9605462\\
  May 1996
\end{tabular}
\end{flushright}
\vskip0.5cm

\title{SIGNATURES OF COLOR-OCTET QUARKONIUM PRODUCTION\footnote{
To appear in the Proceedings of the Second Workshop on Continuous 
Advances in QCD, Minneapolis, U.S.A., March 1996. Research supported 
by the Department of Energy under contract DE-AC03-76SF00515.}}

\author{M.~BENEKE}

\address{Stanford Linear Accelerator Center, Stanford University,
\\ Stanford CA 94309, U.S.A}


\maketitle\abstracts{
I briefly review the nonrelativistic QCD picture of quarkonium 
production and its confrontation with experiment in various 
production processes. 
}

\section{Introduction}
\label{intro}

Quarkonium spectroscopy, decay and production has provided us with 
an interesting place to test our ideas on QCD ever since charmonium was 
discovered in 1974. Yet, the potential of perturbative QCD (PQCD) 
to treat production and decay has been 
fully exploited only recently \cite{BOD95} 
in a development comparable to that of Heavy Quark Effective Theory 
for heavy-light mesons. About the same time, experiments measuring 
quarkonium production at large transverse momentum have confronted 
theorists with surprizingly large cross sections. These observations 
have led to the understanding that fragmentation \cite{BRA93} and 
hadronization of color-octet \cite{BOD92a,BRA95} quark-antiquark 
($Q\bar{Q}$) pairs are essential in the production process. Color-octet 
mechanisms were considered in quarkonium decays already a while 
ago \cite{VOL84}. They were found to solve the problem of infrared divergences 
in P-wave decays in a systematic way \cite{BOD92b}. Taking them into account 
also in S-wave production, where they are not required by perturbative 
consistency in leading order of a nonrelativistic expansion, opens 
the promise of a quantitative description of quarkonium production.

In this talk I briefly summarize the concepts underlying the 
theory of quarkonium production. The subsequent survey 
of the importance of color-octet production concentrates on $J/\psi$ 
production. [Of course, most results on direct $J/\psi$ production 
generalize to $\psi(nS)$ and $\Upsilon(nS)$ states.] Because of space 
limitations, the presentation is sometimes sketchy and for many details 
the reader should consult the original literature, especially to 
appreciate the sources of uncertainty in comparing theory with 
experiment.   

\section{Theory of Quarkonium Production}
\label{theory}

\subsection{Factorization}
\label{factorization}

\begin{figure}[t]
   \vspace{0cm}
   \epsfysize=4.7cm
   \epsfxsize=7cm
   \centerline{\epsffile{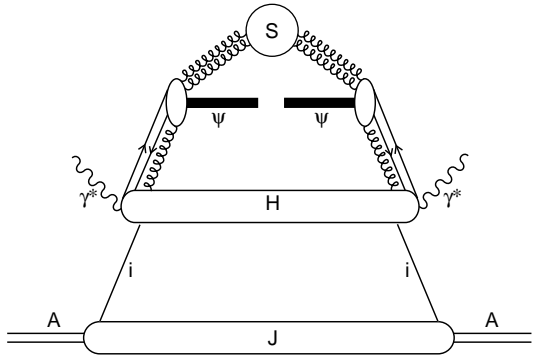}}
   \vspace*{0cm}
\caption{\label{fig} Diagrammatic representation of factorization in 
$\gamma^*+A\to \psi+X$. For the cross section the diagram has to be 
cut.} 
\end{figure}

Inclusive quarkonium production involves two distinct scales. First, 
a heavy quark pair is produced on a distance scale of order $1/m_Q$. 
Then, the quark pair is bound into a quarkonium on a time scale 
of order of the inverse binding energy, $\tau\sim 1/(m_Q v^2)$ (in the 
quarkonium rest frame), where $v$ is the typical velocity of the 
bound quarks. We assume that $v^2$ is small (but do not assume that 
the binding force is Coulombic). The creation process can be computed 
in PQCD and is insensitive to the details of the bound state. The 
binding process can not be computed perturbatively, but long-wavelength 
gluons responsible for binding do not resolve the short-distance production 
process. The factorization hypothesis for quarkonium production \cite{BOD95} 
states that the quarkonium production cross section can be written as a 
sum of short-distance coefficients that describe the creation of 
a $Q\bar{Q}$ pair in a state $n$ multiplied by a process-independent 
matrix element that parameterizes the `hadronization' of the $Q\bar{Q}$ 
state $n$ into a quarkonium $\psi$ plus light hadrons with energies 
of order $m_Q v^2$ in the quarkonium rest frame. 
Consequently, in a hadron-hadron collision 
$A+B\to \psi+X$, the differential cross section is given by
\be
d\sigma=\sum_{i,j}\int d x_1 d x_2\,f_{i/A}(x_1) f_{j/B}(x_2)\, 
\sum_n d\hat{\sigma}_{i+j\to Q\bar{Q}[n]}\,\langle{\cal O}^\psi_n\rangle,
\label{fact}
\ee
where $f_{i/A}$ denotes the parton distribution function. The factorization 
formula is diagrammatically represented in Fig.~\ref{fig} for deeply 
inelastic scattering. The upper part of the diagram represents the 
matrix element $\langle\cal{O}^\psi_n\rangle$. The hard part $H$ is 
connected to this matrix element by $Q\bar{Q}$ lines plus additional 
lines, if ${\cal O}^\psi_n$ contains more fields. Factorization entails 
that soft gluons connecting $S$, $H$ and the remnant jet $J$ cancel up 
to `higher twist' effects in $\Lambda$, where $\Lambda$ represents the 
QCD low-energy scale. Each element in Eq.~\ref{fact} depends 
on a factorization scale. Since gluon emission changes the $Q\bar{Q}$ state 
$n$, a change of factorization scale reshuffles contributions 
between different terms in the sum over $n$ and only the sum is 
physical. The short-distance cross sections $d\hat{\sigma}$ are computed 
by familiar matching: One first calculates the cross section for a 
perturbative $Q\bar{Q}$ state and then subtracts the matrix elements 
computed in this state.

A heuristic argument for factorization can be given starting from the 
infrared finiteness of open heavy quark production. The above 
short-distance cross sections are obtained by expanding the amplitude 
squared for open heavy quark production in the relative three-momentum 
of the quarks and by taking projections on color and angular momentum 
states. Since soft gluon emission takes one from one state to another, 
each projection separately is not infrared safe. However, by construction 
the sum in $n$ runs over all states, 
so that infrared sensitive contributions can always 
be absorbed in some matrix element. In this sense, the sum over all 
intermediate $Q\bar{Q}$ states restores the inclusiveness of open heavy 
quark production. It is also worth noting that Eq.~\ref{fact} 
is valid, up to higher twist effects, 
even if the quarkonium is produced predominantly at small transverse momentum 
with respect to the beam axis, provided one integrates over 
all $p_t$. The transverse momentum distribution is not described by 
Eq.~\ref{fact} unless $p_t\gg\Lambda$. A point of concern, however, is, 
that higher twist effects might be of order $\Lambda/(m_Q v^2)$, 
when a soft gluon from a remnant jet builds the higher Fock state 
$Q\bar{Q}g$ together with the quark pair. 
If these terms do not cancel, Eq.~\ref{fact} 
would be quantitative in hadro-production of quarkonia at low $p_t$ only 
for asymptotically large quark masses, since $\Lambda/(m_Q v^2)\sim 1$ 
both for charmonium and bottomonium.

\subsection{NRQCD and Velocity Scaling}

The matrix elements $\langle {\cal O}^\psi_n\rangle$ contain all 
interactions of the nonrelativistic $Q\bar{Q}$ pair (in its rest frame) 
with degrees of freedom with low momentum compared to $m_Q$. These 
interactions are accurately described by an effective field theory 
called nonrelativistic QCD (NRQCD). The matrix elements are defined 
in NRQCD as \cite{BOD95}
\be
\label{matrix}
\langle {\cal O}^H_n\rangle = 
\sum_X\sum_\lambda\,\langle 0|\,\chi^\dagger {\cal \kappa}_n\psi\,
|H(\lambda)X
\rangle\langle H(\lambda)X|\,\psi^\dagger {\cal \kappa}^\prime_n\chi\,
|0\rangle,
\ee
where the sum is over all polarizations $\lambda$ and light hadrons $X$, 
and $\psi$, $\chi$ are two-spinor fields. [Matrix elements 
with additional gluon fields 
are suppressed in $v^2$ and will not be considered.] Typically, 
the `kernels' 
${\cal \kappa}_n$, ${\cal \kappa}_n^\prime$ specify the color, spin and 
orbital angular momentum state of the quark-antiquark pair. 

Without an additional 
organizing principle, there would be too many matrix elements 
to make the theory predictive. The NRQCD Lagrangian informs us 
about the coupling of soft gluons to the $Q\bar{Q}$ pair. In particular, 
spin symmetry holds to leading order in $v^2$ and reduces the number 
of independent matrix elements considerably. There is no flavor symmetry 
as in Heavy Quark Effective Theory, because the kinetic energy term is 
part of the leading order Lagrangian. Quarkonium spectroscopy tells us 
that the kinetic energy is approximately constant in the range of 
reduced masses that comprises $c\bar{c}$ and $b\bar{b}$ states. Thus, 
$v^2\sim 1/m_Q$ in the range of interest, while for very large quark masses 
$v^2\sim 1/\ln^2 m_Q$. Furthermore, the overlap of the final state 
$H X$ with a $Q\bar{Q}$ state $n$ can be estimated from a multipole 
expansion, which allows us to drop some operators that acquire an 
additional suppression compared to the scaling in $v^2$ of the kernels 
themselves. The resulting `velocity scaling rules' are summarized in 
Refs.~\cite{BOD95,LEP92}. The double expansion of a quarkonium 
production cross section in $\alpha_s$ and $v^2$ (neglecting higher 
twist corrections) is now complete.

As a result two matrix elements, $\langle {\cal O}_1^{\chi_{c0}}({}^3P_0)
\rangle$ and $\langle {\cal O}_8^{\chi_{c0}}({}^3S_1)\rangle$ are needed 
to describe the production of all three P-wave states at leading order 
in $v^2$. The notation refers to the kernels in Eq.~\ref{matrix}: 
The subscript denotes the color state and the angular momentum state is 
written in spectroscopic notation. The first matrix element reduces, to 
leading order in $v^2$, to the familiar derivative of the 
wavefunction at the origin. The 
second color-octet term absorbs the IR senstive regions that would 
otherwise appear in the short-distance cross section \cite{BOD92a,BOD92b}. 
At leading order in $v^2$, $J/\psi$ production is described by the single 
parameter $\langle {\cal O}_1^{J/\psi}({}^3S_1)\rangle$. Because of charge 
conjugation, the gluon-gluon fusion short-distance cross section which 
multiplies this matrix element is suppressed by $\alpha_s$ and is 
proportional to $\alpha_s^3$. Since $v^2\sim 0.25-0.3$ is not very small for 
charmonium, higher order corrections in $v^2$ can be important, if they 
arise at lower order in $\alpha_s$. According to the velocity scaling 
rules three color-octet matrix elements -- 
$\langle {\cal O}_8^{J/\psi}({}^3S_1)\rangle$, 
$\langle {\cal O}_8^{J/\psi}({}^1S_0)\rangle$, 
$\langle {\cal O}_8^{J/\psi}({}^3P_0)\rangle$ -- contribute at order 
$\alpha_s^2 v^4$ (powers of $v^2$ are counted relative to the leading 
order contribution), so that $J/\psi$ production is described by four 
nonperturbative parameters. Contributions of order $\alpha_s^3  v^2$ 
also exist and can be important in specific regions of phase space. 
Because $\chi_{c1}$ production at low tranverse momentum is also 
suppressed by $\alpha_s$ compared to $\chi_{c0}$ and $\chi_{c2}$, 
higher order corrections in $v^2$ would also be important for 
$\chi_{c1}$ production.

\subsection{Fragmentation}

We now consider the transverse momentum distribution $d\sigma/d p_t^2$. 
The leading order contributions (in $\alpha_s$ and $v^2$) decrease as 
$1/p_t^6$ (P-waves) or $1/p_t^8$ (S-waves), when $p_t$ is large compared 
to $2 m_Q$. This steep decrease is a penalty for preforming the quarkonium 
state at very small distances $1/p_t$ rather than $1/m_Q$ and is not what 
would be expected from a high-energy cross section in QCD. For 
$A+B\to H+X$, where $H$ can be any hadron, we expect scaling: When the 
cms energy and $p_t$ are large compared to all hadron masses, the 
short-distance cross section $d\hat{\sigma}/dp_t^2$ scales as 
$1/p_t^4$ on dimensional grounds, modulo logarithms of $p_t$ and higher 
twist corrections of order $m_H/p_t$ and $\Lambda/p_t$. Moreover, the 
leading twist cross section can be written as a convolution of distribution 
functions, a short-distance cross section and a fragmentation function. 
From their $p_t$-behaviour we deduce that the leading order (in $\alpha_s$ 
and $v^2$) $J/\psi$ production mechanisms are higher twist at large $p_t$ 
(but calculable, since $m_H$ is large compared to $\Lambda$).  
In general, fragmentation functions remain uncalculable. If $H$ is a 
quarkonium, however, the dependence on the energy fraction 
$z$ can be calculated \cite{BRA93,MA94}, because the quark mass 
$m_Q\gg\Lambda$ provides another large mass scale. Thus, a parton fragments 
first into a $Q\bar{Q}$ pair, which subsequently hadronizes. 
Since the hadronization 
of the $Q\bar{Q}$ pair takes place by emission of gluons with momenta 
of order $m_Q v^2$ in the quarkonium rest frame, 
the energy fraction of the quarkonium relative to 
the fragmenting parton, differs from that of the $Q\bar{Q}$ pair 
only by an 
amount $\delta z\sim v^2\ll 1$. As a result, the fragmentation functions 
are expressed as a sum over perturbatively calculable, 
$z$-dependent coefficient functions that 
describe the fragmentation process $i\to Q\bar{Q}[n]$ multiplied by the 
same $z$-independent `hadronization matrix elements' encountered 
earlier.

\renewcommand{\arraystretch}{1.5}
\begin{table}[t]
\caption{Parametric dependence of various $J/\psi$ production mechanisms 
in hadron-hadron collisions. 
\label{tab1}}
\vspace{0.4cm}
\begin{center}
\begin{tabular}{|c|c|c|}
\hline
& color singlet & color octet \\ \hline
$p_t\sim 0$ & $\alpha_s^3$ & $\alpha_s^2 v^4$ \\ \hline
$p_t \gg 2 m_Q$ & $\alpha_s^5 \left(p_t^2/(4 m_Q^2)\right)^2$ & 
$\alpha_s^3 v^4 \left(p_t^2/(4 m_Q^2)\right)^2$\\
\hline
\end{tabular}
\end{center}
\end{table}
At large $p_t$ quarkonium production depends on three small parameters: 
$\alpha_s/\pi\sim 0.1$, $v^2\sim 0.25-0.3$, $4 m_Q^2/p_t^2$ (numbers for 
charmonium). The leading contributions for octet vs. singlet and 
fragmentation vs. non-fragmentation production of $J/\psi$ are shown 
in Tab.~\ref{tab1}. The scaling with $v^2$ and $4 m_c^2/p_t^2$ is 
measured relative to the color-singlet non-fragmentation term. We see 
that at $p_t\sim 10\,$GeV gluon fragmentation into a color-octet 
quark pair dominates \cite{BRA95} all other mechanisms by at least a 
factor ten.

\subsection{Quarkonium Polarization}

Quarkonium polarization in the NRQCD formalism has been considered 
in Refs.~\cite{BR1,MA95,BR2,BRA96a}. A novel feature as compared to 
unobserved polarization is that the short-distance cross sections 
can not be determined from a matching calculation that involves 
only amplitudes squared of $Q\bar{Q}$ states with definite angular 
momentum. The final quarkonium state can be reached through 
quark-antiquark pairs in various spin and orbital angular momentum 
states, which are coherently produced, so that interference 
between different intermediate states occurs \cite{BR1,BR2,BRA96a}.  
Because of parity and charge conjugation symmetry, intermediate states with 
different spin $S$ and angular momentum $L$ can not 
interfere, so that for $J/\psi$ production interference occurs for 
intermediate ${}^3 P_J$-states only. The interference terms are crucial 
to obtain a factorized expression for fragmentation functions into 
polarized $\psi(nS)$ states \cite{BR1}.

The amplitude projections that determine the short-distance cross 
section can be found as follows. The cross section can be written 
as (see Fig.~\ref{fig})
\be
\label{factor}
\sigma^{(\lambda)} \sim H_{ai;bj}\cdot S_{ai;bj}^{(\lambda)}\,,
\ee
where the indices $ij$ and 
$ab$ refer to spin and orbital angular momentum 
in a Cartesian basis $L_a S_i$ 
($a,i=1,2,3=x,y,z$). For $Q\bar{Q}$ with spin and orbital angular 
momentum one, the soft part is given by  
\begin{equation}
S_{ai;bj}^{(\lambda)}=\sum_X
\langle 0|\chi^\dagger\sigma_i T^A\!\left(\!-\frac{i}{2}\! 
\stackrel{\leftrightarrow}{D}_a\!\right)\! 
\psi\,|J/\psi(\lambda) X\rangle\langle J/\psi(\lambda) X|
\,\psi^\dagger\sigma_j T^A\!\left(\!-\frac{i}{2}\! 
\stackrel{\leftrightarrow}{D}_b\!\right)\!\chi|0\rangle\,,
\end{equation}
where helicity $\lambda$ is fixed. To proceed one writes down the most 
general tensor decomposition compatible with rotational invariance, parity 
and charge conjugation. In general, this introduces a substantial number 
of new nonperturbative parameters. To evaluate the matrix element at 
leading order in $v^2$, we may use spin symmetry. Spin symmetry implies that 
the spin of the $J/\psi$ is aligned with the spin of the $c\bar{c}$ pair, 
so $S_{ai;bj}^{(\lambda)}\propto 
{\epsilon^i}^*(\lambda)\epsilon^j(\lambda)$. One then finds
\begin{equation}
\label{decomp}
S_{ai;bj}^{(\lambda)}=
\langle {\cal O}^{J/\psi}_8(^3\!P_0)\rangle\,
\delta_{ab}\,{\epsilon^i}^*(\lambda)\epsilon^j(\lambda)\,.
\end{equation}
All other possible Lorentz structures are suppressed by $v^2$. Moreover, 
the single surviving structure is related to a matrix element that appeared 
already in unpolarized production. This result is 
general \cite{BR1,BR2,BRA96a}: At order $\alpha_s^2 v^4$, no new matrix 
elements are required to describe $J/\psi$ polarization.

The decomposition Eq.~\ref{decomp} tells us that to calculate 
the polarized production
rate we should project the hard scattering amplitude onto states
with definite $S_z=\lambda$ and $L_z$, square the 
amplitude, and then sum over 
$L_z$ ($\sum_{L_z}\epsilon_a(L_z)
\epsilon_b(L_z)=\delta_{ab}$  in the rest frame). The 
soft part is diagonal in the $L_z S_z$ basis. It is straightforward 
to transform to the more conventional $J J_z$ basis. Since 
$J_z=L_z+S_z$, there is no interference between intermediate states 
with different $J_z$. We write, with obvious 
notation, 
\begin{equation}
\label{factornew}
\sigma^{(\lambda)} \sim \sum_{J J_z;J' J_z^\prime} 
H_{J J_z;J' J_z^\prime}\cdot S_{J J_z;J' J_z^\prime}^{(\lambda)}\,,
\end{equation}
and obtain, using Eq.~\ref{decomp},  
\begin{equation}
S_{J J_z;J' J_z^\prime}^{(\lambda)} =
\langle {\cal O}_8^{J/\psi} ({}^3 P_0)\rangle 
\sum_M \langle 1M;1\lambda|J J_z\rangle\langle J' J_z^\prime|1 M;1 \lambda
\rangle\,,
\end{equation}
which is diagonal in $(J J_z)(J' J_z^\prime)$ only after 
summation over $\lambda$ (unpolarized production). In general, the 
off-diagonal matrix elements cause interference of the following 
$J J_z$ states: $00$ with $20$, $11$ with $21$ and $1(-1)$ with $2 (-1)$. 
This particular pattern of interference is a consequence of spin 
symmetry. In general, all states interfere, and diagonality in the 
$L_z S_z$ basis is also lost.

\section{Confrontation with Experiment}

The interest in color-octet production mechanisms was ignited by the 
possibility to explain the large $\psi'$ production cross section 
at the Tevatron by gluon fragmentation into a color-octet $c\bar{c}$ 
pair. Over the past year most $J/\psi$ production processes have been 
reanalyzed with color-octet mechanisms taken into account. In this section 
I intend to give a short summary. References to experimental results 
can be found in the quoted papers. The letter `$\psi$' denotes 
$J/\psi$ and $\psi'$ collectively.

\subsection{Large $p_t$}

Quarkonium production at large transverse momentum with respect to 
the beam axis can be measured at colliders, most recently in $p\bar{p}$ 
collisions at the Tevatron, where the indirect contribution from 
$B$ decays can be removed. The accessible range of $p_t$ is 
$5\,\mbox{GeV}<p_t<20\,\mbox{GeV}$. In this range fragmentation dominates. 
The $J/\psi$ production data, which includes feed-down from P-wave 
states could be described \cite{CAC94} - 
within theoretical and experimental errors - 
by fragmentation into P-wave states together with 
color-singlet gluon fragmentation into $J/\psi$. 
The $\psi'$ cross section, however, 
is under-predicted by a factor 30 by color singlet fragmentation. 
The same deficit was found for direct $J/\psi$ production, after the 
contributions from higher charmonium states could be separated. 
The dominant S-wave production mechanism was still missed.

The discrepancy can be explained \cite{BRA95,CAC95} 
by taking into account that a gluon 
can fragment into a color octet $c\bar{c}$ pair in a ${}^3 S_1$ state, 
which decays into a $\psi$ state by emission of two gluons with 
momenta of order $m_c v^2$ in the 
quarkonium rest frame. The suppression factor $v^4$ is more than 
compensated by the enhanced short-distance coefficient, see Tab.~\ref{tab1} 
and Sect.~2.3. At lowest order in $\alpha_s$ only 
$\langle {\cal O}_8^\psi({}^3 S_1)\rangle$ is probed by gluon 
fragmentation and can be determined by fitting the Tevatron data. 
The resulting $\langle {\cal O}_8^\psi({}^3 S_1)\rangle/
\langle {\cal O}_1^\psi({}^3 S_1)\rangle\sim 1/100$ is actually 
smaller than $v^4\sim 1/10$ as expected from velocity scaling. Apart 
from numerical factors in the short-distance coefficients, this could 
partly be due to the fact that the emission of soft gluons is 
kinematically not accounted for in the octet fragmentation function. 
Soft gluon emission softens the fragmentation function by smearing the 
delta-function over a region $\delta z\sim v^2$ in longitudinal 
momentum fraction. Since the short-distance cross section falls 
off like $1/p_t^4$, one roughly probes the fourth moment of the 
fragmentation function. A softer fragmentation function would then 
require a larger matrix element to fit the data.

The theoretical predicition was extended to moderate $p_t\sim 2 m_c$ 
(or slightly larger) in Ref.~\cite{CHO95}, where contributions 
suppressed by $4 m_c^2/p_t^2$ at large $p_t$ have been kept. In 
this intermediate region, all three color-octet matrix elements 
relevant for $\psi$ production are equally significant, but should 
be suppressed by $v^4$ compared to the color-singlet cross section, 
since no compensating factors of $\pi/\alpha_s$ are at work. 
Surprizingly, due to numerical enhancements of the amplitudes, the 
color-octet contributions still dominate and seem to be required 
to describe the data. The best fit yields a value of 
$\langle {\cal O}_8^{J/\psi}({}^3 S_1)\rangle$ 
which is more than a factor two smaller 
than the one extracted \cite{CAC95} from large $p_t$, while the 
fitted combination of 
$\langle {\cal O}_8^{J/\psi}({}^3 P_0)\rangle$ and 
$\langle {\cal O}_8^{J/\psi}({}^1 S_0)\rangle$ is rather large, 
especially in comparison with the fits for $\psi'$. Whether this 
difference between $J/\psi$ and $\psi'$ is an artefact or physical 
effect remains to be decided.

As noted in Ref.~\cite{WIS95}, the above scenario implies, that 
the directly produced $\psi$ mesons are almost completely transversely 
polarized at large $p_t$. The fragmenting gluon is transversely 
polarized and transfers its polarization to the $c\bar{c}$ pair in 
the ${}^3 S_1$ state. Because of spin symmetry, the transverse 
polarization stays intact in the emission of two soft gluons. Polarization 
would be measured in the angular distribution of leptonic $\psi$ 
decay, $d\Gamma/d\cos \theta \propto 1+\alpha \cos^2\theta$, where 
$\theta$ denotes the angle between the lepton three-momentum in the 
$\psi$ rest frame and the $\psi$ three-momentum in the lab frame. 
Higher order corrections to the fragmentation function can 
result in longitudinally polarized $\psi$, but were found to be 
rather small \cite{BR1}. For 
$\langle {\cal O}_8^{\psi}({}^3 P_0)\rangle/ 
(m_c^2 \langle {\cal O}_8^{\psi}({}^3 S_1)\rangle) < 2$, one still has 
$\alpha>0.64$. Spin-symmetry breaking corrections are suppressed by 
$v^4$ and introduce an uncertainty of $0.1$ in $\alpha$. These 
corrections to $\alpha=1$ persist at large $p_t$. For moderate 
$p_t$ corrections that vanish as $4 m_c^2/p_t^2$ exist, but have 
not yet been calculated. Since other corrections are small, these 
could well dominate over most of the $p_t$-range accessible at the 
Tevatron. So far, the angular distribution remains unmeasured. There is 
a price in statistics to pay, but understanding the angular 
dependence of the detector acceptance is also non-trivial. Ideally, 
a measurement of the $p_t$-dependence of $\alpha$ will provide 
a decisive self-consistency test of the octet production 
picture.  

At HERA large-$p_t$ production of $J/\psi$ is probed in 
$\gamma p$ collisions. In this case gluon fragmentation plays no 
role, since at leading order in $\alpha_s$ only $\gamma q$ 
collisions contribute to this mechanism. 
Color singlet charm fragmentation, 
induced by photon-gluon fusion, dominates 
\cite{GOD96} at large $p_t$.

\subsection{Small $p_t$: Fixed Target Experiments}

Fixed target experiments have for a long time been the most 
profuse source of quarkonium production data. Since no cut on 
$p_t$ is usually imposed, the production cross section is dominated 
by quarkonia with transverse momenta of about $1\,$GeV. The 
cms energies range up to $\sqrt{s}=40\,$GeV. The most recent 
comparisons of color-singlet production mechanisms with data 
are documented in Refs.~\cite{SCH94,VAE95}. Summarizing the 
conclusions of Ref.~\cite{VAE95}, the color singlet mechanisms 
(a) do not account for the over-all normalization of the total 
cross section very well, (b) yield too low a fraction of directly 
produced $J/\psi$, (c) transverse rather than no polarization 
of $J/\psi$ and $\psi$ and (d) predict far too few $\chi_{c1}$ 
states in comparison to $\chi_{c2}$. With these failures in mind, 
color-octet mechanisms have been analyzed \cite{BR2,TAN95,GUP96}. 
[For reasons explained in Ref.~\cite{BR2}, there are substantial 
differences in numerics and conclusions between these 
three analyses. The subsequent presentation adheres to 
Ref.~\cite{BR2}.] The results are 
not entirely encouraging, but not uninteresting either.

Because of charge conjugation, color-singlet $\psi$ production 
is suppressed by $\alpha_s$. Consequently, octet mechanisms scale 
as $(\pi/\alpha_s)\,v^4\sim 1$, see Tab.~\ref{tab1}. Since the 
color-singlet amplitude vanishes at $c\bar{c}$ threshold while 
the octet amplitude does not, and since the cross section is 
enhanced at threshold by the $x$-behaviour of the gluon distribution, 
color-octet production actually dominates the cross section. 
At leading order, the combination 
$\Delta_8(\psi)=\langle {\cal O}_8^{\psi}({}^1 S_0)\rangle+7/m_c^2 
\langle {\cal O}_8^{\psi}({}^3 P_0)\rangle$ is probed. At this 
order $\langle {\cal O}_8^{\psi}({}^3 S_1)\rangle$ arises only 
in $q\bar{q}$ annihilation, which is numerically small compared to 
gluon-gluon fusion in the energy range considered. 

Starting with total unpolarized cross sections, one has three 
observables - $\sigma_{J/\psi}$, $\sigma_{\psi'}$ and the 
direct $J/\psi$ cross section $\sigma_{J/\psi}^{dir}$, which does 
not include feed-down from higher charmonium states - and two 
octet hadronization parameters - $\Delta_8(J/\psi)$, 
$\Delta_8(\psi')$ - to fit. [If these were accurately known from 
other processes, no free parameter would remain.] Such a fit is possible 
with matrix elements consistent with their expected size from velocity 
scaling. The energy dependence of the total cross section is in 
agreement with the data, although this can not be considered very 
significant. A fit to $\sigma_{J/\psi}$ and $\sigma_{\psi'}$ 
then predicts a direct $J/\psi$ production fraction of about 
$60\%$, in good agreement with experiment. With respect to the 
failures (a) and (b) above, this could be considered as strong 
evidence for important color-octet contributions in direct 
$\psi$ production. The uncertainties, 
however, are appreciable. For example, in the above fit, 
the color-singlet matrix elements were taken from Buchm\"uller-Tye 
wavefunctions at the origin and not considered as free parameters. 
The over-all normalization depends strongly on the charm quark mass, 
although the direct production fraction does not. The short-distance 
coefficients have been expressed in terms of the charm quark mass 
rather than quarkonium masses. This is conceptually preferred by the 
factorization formalism and numerically quite important. 

Even disregarding these uncertainties, the above picture can not be 
considered as complete. Defining a parameter $\alpha$ for the 
angular distribution as before, where $\theta$ is now the angle 
between the three-momentum vector of the positively charged muon and 
the beam axis in the quarkonium rest frame, one finds 
$0.15 < \alpha < 0.44$ for $\psi'$ production at 
$\sqrt{s}=21.8\,$GeV and $0.31 < \alpha < 0.63$ for $J/\psi$ production 
at $\sqrt{s}=15.3\,$GeV. [Since the energy dependence is mild, these 
numbers can be used with little error at higher cms energies.] These 
estimates could be made more precise, if the matrix elements 
$\langle {\cal O}_8^{J/\psi}({}^3 P_0)\rangle$ and 
$\langle {\cal O}_8^{J/\psi}({}^1 S_0)\rangle$ were individually 
known. The degree of transverse polarization is higher in $J/\psi$ 
production, because the indirect contribution from $\chi_{c2}$ decays 
yields a purely transversely polarized component to the 
cross section. These estimates 
should be compared with the measurement of no visible polarization, 
$\alpha\approx 0$, both for $\psi'$ and $J/\psi$. 

At leading order in $v^2$, one predicts a $\chi_{c1}:\chi_{c2}$ 
production ratio (weighted by their branching fractions to decay into 
$J/\psi$) of $1:7$, mainly because $\chi_{c1}$ production is 
suppressed in $\alpha_s$. As for direct $J/\psi$ production, higher 
order corrections in $v^2$ can be expected to dominate $\chi_{c1}$ 
production. Although they have not yet been analyzed, they could raise 
the above ratio to about $1:3$, still far below the observed ratio 
$(1.4\pm0.4):1$. Since the $\chi_{c1}:\chi_{c2}$ ratio has been 
measured in only one experiment \cite{LEM82}, an independent confirmation 
of this discrepancy would be welcome.

The discrepancy between predictions and data on polarization might 
be due to numerically large spin symmetry breaking corrections, which 
would lead to less transverse polarization. One 
could also suspect that the experiments have not yet fully accounted 
for all systematic errors. In the E672/E706 experiment (the only 
one that provides this piece of information \cite{GRI96}), 
for instance, the 
acceptance and efficiency varies strongly with angle, while at the 
same time, the acceptance/efficiency curve was determined from a 
Monte Carlo sample of unpolarized quarkonia. However, together with the 
problematic $\chi_{c1}:\chi_{c2}$ ratio, it seems more likely that 
the interactions of a color octet $c\bar{c}$ pair with 
soft gluons as it traverses the target are not understood. 
Although of higher twist in $\Lambda/m_c$, such effects must be 
sizeable as illustrated by the large nuclear dependence of the total 
$\psi$ cross sections, which is dynamically also not understood. 
Since quarkonium formation occurs only after 
the $c\bar{c}$ pair left the nucleus, it appears as if by the 
time the quarkonium is formed, the $c\bar{c}$ pair has lost `memory' 
that it was produced from two on-shell (transverse) gluons, as assumed 
in leading twist. But even if higher twist effects are sizeable, the 
polarization problem seems hard to solve, since the 
coupling of soft gluons to the $c\bar{c}$ pair would be 
subject to spin symmetry and the counting rules for multipole 
transitions, so that transverse polarization is again hard to 
avoid. No higher twist mechanism has yet been shown to produce 
predominantly longitudinal polarization in the central $x_F$ region 
that dominates the cross section. To clarify the higher twist nature 
of the discrepancies, analogous measurements for 
$\Upsilon$ production would be desirable. As another check, it would 
be interesting to know to 
what extent $\chi_{c2}$ is produced with helicity 
$\pm 2$ as predicted at leading twist (with small corrections).

Photoproduction of charmonia in fixed target experiments (or HERA, 
in the low-$p_t$ region) is theoretically rather similar to fixed target 
hadroproduction as far as the underlying diagrams and the probed 
matrix element $\Delta_8(J/\psi)$ is concerned. A comparison of 
octet mechanisms with data was undertaken both in the elastic 
region \cite{CAC96,AMU96,KO96} $z>0.95$ and the inelastic 
domain \cite{CAC96,KO96} $z<0.9$, where $z=p\cdot k_\psi/p\cdot k_\gamma$ 
is the energy fraction $E_\psi/E_\gamma$ in the proton rest frame. In 
the elastic region color-octet contributions are enhanced by 
$\pi/\alpha_s$ compared to color-singlet contributions, as in 
fixed target hadroproduction. The extracted matrix element 
$\Delta_8(J/\psi)$ is consistent with the extraction from hadroproduction, 
but substantially smaller than what would have been expected from 
the fit to the Tevatron data in the moderate $p_t$ domain discussed 
earlier. While the consistency with hadroproduction is reassuring, 
the situation with higher twist effects is even more delicate in 
the large-$z$ region than in hadroproduction. As octet mechansims, 
`diffractive' quarkonium production is also centered at $z\approx 1$ 
and unsuppressed if not dominant in the endpoint region. In addition, 
the restriction $z>0.95$ might not allow the process to be sufficiently 
inclusive, as is necessary in the NRQCD approach. With these 
remarks of caution in mind, the discrepancy in color-octet matrix 
elements between `high' energy (Tevatron) and `low' energy 
fits should probably not be over-interpreted. It is likely that 
energy dependent higher-order radiative corrections (such as small-$x$ 
effects) would, once taken into account, lower the matrix elements 
extracted from the Tevatron fits. Such a reduction would be welcome 
to explain a factor two discrepancy in comparison with large-$p_t$ UA1 
data \cite{CAC95} and the $J/\psi$ branching fraction in $B$ decays, 
see below.

In the inelastic domain $z<0.9$, color-octet processes 
are parametrically suppressed by $v^4$, but, due to enhancements 
in the amplitudes, 
numerically of the same order as color-singlet processes. When 
$z$ approaches unity, the color-octet contributions diverge as 
expected from a partonic process with a lowest order term 
proportional to $\delta(1-z)$. This divergence indicates that the 
partonic prediction must be considered as a distribution. When  
smearing is applied, together with the constraints on the 
matrix elements 
$\langle {\cal O}_8^{J/\psi}({}^3 P_0)\rangle$ and 
$\langle {\cal O}_8^{J/\psi}({}^1 S_0)\rangle$ from the elastic 
peak, the presence of color-octet contributions does not appear to be 
in conflict with the $J/\psi$ energy distribution in fixed target 
experiments or at HERA.
 
\subsection{$e^+ e^-$ Annihilation and $Z^0$ Decay}

Quarkonium production in $e^+ e^-$ annihilation shows much of the 
variety of production mechanisms in hadro- and photoproduction, while 
eliminating the uncertainties related to hadrons in the initial 
state. On the other hand, the cross sections are not large and there 
is not much data to compare with. [For instance, out of 3.6 million 
hadronic $Z^0$ decays, OPAL has found 24 prompt $J/\psi$.]

 The situation when $\sqrt{s}$ is not much larger than $2 m_c$, 
relevant at CLEO, has been investigated in Ref.~\cite{CHEN}. The 
parametric dependence of color-octet contributions to the 
charmonium energy spectrum is analogous to the energy spectrum 
in photoproduction. At leading order in $\alpha_s$, the 
$c\bar{c}$ pair can only be produced in an octet state, which should 
be seen as a jet containing a charmonium, recoiling against a gluon 
jet. The energy distribution is peaked at the endpoint 
$E_{max}=(s+M_\psi^2)/(2 \sqrt{s})$, a delta-function smeared 
over a region $v^2 E_{max}$. Away from the endpoint, color-singlet 
contributions are more important. In addition to the energy spectrum, 
the $J/\psi$ angular distribution leads to a striking signature 
of octet production: Close to the endpoint of the energy spectrum, 
color-singlet production favors $J/\psi$ production perpendicular 
to the beam axis, while octet mechanisms produce $J/\psi$ 
predominantly along the beam axis. An accurate prediction again 
requires the values of the matrix elements  
$\langle {\cal O}_8^{J/\psi}({}^3 P_0)\rangle$ and 
$\langle {\cal O}_8^{J/\psi}({}^1 S_0)\rangle$. 

 When $s\gg 4 m_c^2$, as relevant for $Z^0$ decay at LEP, the production 
patterns change. The ratio $4 m_c^2/M_Z^2$ is now an important 
parameter, just as $4 m_c^2/p_t^2$ at large $p_t$. The dominant 
production mechanisms are of fragmentation-type. Although suppressed 
by $v^4$ compared to color-singlet charm fragmentation, gluon 
fragmentation into a color-octet $c\bar{c}$ pair in a ${}^3 S_1$ 
state wins \cite{CHE95,CHO} over charm fragmentation by a factor 
of approximately three, depending on the choice of values for the 
color-singlet and octet matrix elements. 
Gluon fragmentation benefits from larger 
color and flavor factors (all $Z^0\to q\bar{q}$ initiate gluon 
fragmentation), but most importantly from a double logarithmic 
enhancement in $4 m_c^2/m_Z^2$ from the phase space region, where 
the fragmenting gluon is collinear to and softer than the primary 
quark. Consequently, color-octet gluon fragmentation could be 
discriminated from charm fragmentation that yields more energetic 
charmonia \cite{CHE95}. 

Present data from LEP \cite{OPAL} have too low statistics 
to idenitfy a gluon 
fragmentation contribution unambiguously. The central value for 
the prompt $J/\psi$ branching fraction 
$\mbox{Br}(Z^0\to J/\psi\,\mbox{prompt}+X)=(1.9\pm 0.7\pm 0.5\pm 0.5)
\cdot 10^{-4}$ however, is about a factor of two larger than the sum 
of all color-singlet contributions. [Such statements are not quite 
exact, since the experimental efficencies and thus results 
depend on the assumptions on the underlying production mechanism.  
The largest experimental error arises from the determination of the 
prompt production fraction.] The measured energy spectrum is neither 
peaked towards low energies nor towards large energies. Closing the 
eyes on the present errors, one would suspect a roughly equal 
contribution from charm and color-octet gluon fragmentation.

\subsection{Bottom Decay}

Color-octet mechanisms have been first considered for the 
inclusive $\chi_{cJ}$ yield in $B$ meson decays \cite{BOD92a}. In 
leading logarithmic approximation, color singlet production 
is proportional to $(2 C_+-C_-)^2\approx 0.16$, while 
octet production is proportional to $(C_+ + C_-)^2\approx 4.9$, where 
$C_+$ and $C_-$ are related to the Wilson coefficients of the 
current-current operators in the $\Delta B=1$ effective Hamiltonian. 
This numerical enhancement in the Wilson 
coefficients allows color-octet contributions to compete with 
color-singlet contributions, even though the first are suppressed 
by $v^4$ in the matrix elements \cite{KO96,KO2}. The measured 
branching fraction $\mbox{Br}(B\to J/\psi+X)=(0.80\pm 0.08)\%$ 
allows us, in principle, to put a bound on the least known 
matrix elements 
$\langle {\cal O}_8^{J/\psi}({}^3 P_0)\rangle$ and 
$\langle {\cal O}_8^{J/\psi}({}^1 S_0)\rangle$. But the cancellations 
in the combination $(2 C_+-C_-)^2$ also render its numerical 
value highly sensitive to the choice of scale in the Wilson 
coefficients. This, together with the uncertainties in the 
wavefunction at the origin, leave an uncertainty of about a factor 
four in the color-singlet contribution. However, even with no 
color-singlet contribution at all and 
no contribution from $\langle {\cal O}_8^{J/\psi}({}^3 S_1)\rangle$, 
the Tevatron fit \cite{CHO95} of 
$\langle {\cal O}_8^{J/\psi}({}^1 S_0)\rangle+3/m_c^2
\langle {\cal O}_8^{J/\psi}({}^3 P_0)\rangle\sim 6.5\cdot 10^{-2}$ 
would overestimate the branching fraction by about $50\%$. 
Thus, $B$ decays favor 
the smaller values of these matrix elements inferred from 
fixed target hadro- and photoproduction.

Bottomonium decay provides another $J/\psi$ production process. The 
two most important production mechanisms are \cite{TRO94,CHEUNG} 
$\Upsilon\to\chi_{cJ}+X$, dominated by octet $\chi_{cJ}$ production, 
followed by radiative $\chi_{cJ}$ decay, and 
$\Upsilon\to ggg^*$, followed by $g^*\to c\bar{c}[{}^3 S_1^{(8)}]$. 
The second mechanism is similar to color-octet gluon fragmentation, 
although the $\Upsilon$ mass is not large enough to justify the 
fragmentation approximation. The branching fractions are estimated 
to be $7\cdot 10^{-5}$ and $2.5\cdot 10^{-4}$, respectively, 
using the larger 
value $\langle {\cal O}_8^{J/\psi}({}^3 S_1)\rangle=0.015\,$GeV$^3$. 
Their sum is only a factor of two below the upper 
bound $6.8\cdot 10^{-4}$ 
from ARGUS and consistent with the CLEO result 
$(1.1\pm 0.4)\cdot 10^{-3}$ within $2\sigma$. 

\section{Conclusion}
\renewcommand{\arraystretch}{1.5}
\begin{table}[t]
\caption{Importance of color octet contributions in various $J/\psi$ 
production processes and the octet matrix elements that could be 
probed in the 
corresponding process. 
\label{tab2}}
\vspace{0.4cm}
\begin{center}
\begin{tabular}{|l|c|c|c|}
\hline
Process & Reference & $\frac{\mbox{\scriptsize octet}}{\mbox{\scriptsize 
singlet}}$ & Matrix elements \\ \hline
$p\bar{p}$, large $p_t$ & Ref.~\cite{BRA95,CAC95} & 30 & 
${}^3 S_1^{(8)}$ \\ \hline
$p\bar{p}$, moderate $p_t$ & Ref.~\cite{CHO95} & 8 & 
${}^3 S_1^{(8)}$, ${}^3 P_0^{(8)}$, ${}^1 S_0^{(8)}$ \\ \hline
Hadroproduction, fixed target & Ref.~\cite{BR2,TAN95,GUP96} & 2-8 &
${}^3 P_0^{(8)}$, ${}^1 S_0^{(8)}$ \\ \hline
Photoproduction, $z>0.95$ & Ref.~\cite{CAC96,AMU96,KO96} & 4 & 
${}^3 P_0^{(8)}$, ${}^1 S_0^{(8)}$ \\ \hline
Photoproduction, $z<0.9$ & Ref.~\cite{CAC96,KO96} & 1 & 
${}^3 P_0^{(8)}$, ${}^1 S_0^{(8)}$ \\ \hline
$Z^0$ decay & Ref.~\cite{CHE95,CHO} & 3 & 
${}^3 S_1^{(8)}$ \\ \hline
$e^+ e^-$ annihilation, large $z$ & Ref.~\cite{CHEN} & 4  & 
${}^3 P_0^{(8)}$, ${}^1 S_0^{(8)}$ \\ \hline
$B$ decay & Ref.~\cite{KO96,KO2} & 3 & 
${}^3 S_1^{(8)}$, ${}^3 P_0^{(8)}$, ${}^1 S_0^{(8)}$ \\ \hline
$\Upsilon$ decay & Ref.~\cite{TRO94,CHEUNG} & 3 & 
${}^3 S_1^{(8)}$ \\ \hline
\end{tabular}
\end{center}
\end{table}

The magnitude of color-octet vs. color-singlet production cross 
sections for the $J/\psi$ production processes discussed 
in Sect.~3 are summarized in 
brief in Tab.~\ref{tab2}, together with the hadronization matrix elements 
probed in the corresponding process. A rather striking observation 
is that color-octet mechanisms dominate every production process. 
So, why are we discussing them only more than twenty years after the 
discovery of charmonium?

For one thing, many accurate experimental results are rather recent. 
Another is that the charm mass is not really large enough to make 
precise predictions. Thus, a dramatic signature such as in large-$p_t$ 
production at the Tevatron was necessary to induce a reanalysis of 
other processes. The numbers in the table should be 
considrered as indicative only  
and are often uncertain by a factor of two or more. 
Even now these uncertainties do 
not allow us to determine all production matrix elements. With the 
correct theoretical framework at hand, further effort, both experimental 
and theoretical, can be undertaken to make the presented 
quarkonium production picture fully 
quantitative. 

\section*{Acknowledgements}
I wish to thank Ira Rothstein for his collaboration 
on the subject discussed in this 
article and Arthur Hebecker for very useful discussions. 
Thanks to both for reading the manuscript.

\end{document}